\documentclass[showpacs]{revtex4}
\usepackage{graphicx,amsmath,amssymb,amsfonts,bbm,latexsym,color,dcolumn,epsf}

\def\be{\begin{equation}}
\def\ee{\end{equation}}
\def\lp{\left(}
\def\rp{\right)}
\def\lb{\left[}
\def\rb{\right]}
\def\om{\omega}

\def\la{\lambda}
\def\ck{\chi_k}

\begin{document}

\title{Backreaction in trans-Planckian cosmology: renormalization, trace anomaly and selfconsistent solutions}
\author{D. L\'opez Nacir \footnote{dnacir@df.uba.ar}}
\author{F. D. Mazzitelli \footnote{fmazzi@df.uba.ar}}
\affiliation{Departamento de F\'\i sica {\it Juan Jos\'e
Giambiagi}, Facultad de Ciencias Exactas y Naturales, UBA, Ciudad
Universitaria, Pabell\' on I, 1428 Buenos Aires, Argentina}

\begin{abstract}
We analyze the semiclassical Einstein equations for quantum scalar
fields satisfying modified dispersion relations. We first discuss
in detail the renormalization procedure based on adiabatic
subtraction and dimensional regularization. We show that, contrary
to what expected from power counting arguments, in $3+1$
dimensions the subtraction involves up to the fourth adiabatic
order even for dispersion relations containing higher powers of
the momentum. Then we analyze the dependence of the trace of the
renormalized energy momentum tensor with  the scale of new
physics, and we recover the usual trace anomaly in the appropriate
limit. We also find selfconsistent de Sitter solutions for
dispersion relations that contain up to the fourth power of the
momentum. Using this particular example, we also discuss the
possibility that the modified dispersion relation can be mimicked
at lower energies by an effective initial state in a theory with
the usual dispersion relation.

\end{abstract}


\maketitle

\section{Introduction}
The expansion of the Universe can act as a cosmological
microscope. Scales of interest today could have been
sub-Planckian at the first stages of inflation, and  it has been
therefore speculated that there could be signatures of
trans-Planckian physics in the evolution of the universe and/or in
the inhomogeneities of the CMBR \cite{bran1,transp1}. It is of
course very difficult to address these issues without knowing the
physics at Planck scale. A plausible approach is to test the
robustness of inflationary predictions under  different changes
that could be attributable to the unknown new physics. For
example, it has been argued that one of the effects could be the
modification of the dispersion relations for the modes of the
quantum fields with momenta larger than a given scale $M_c$, as
suggested by loop quantum gravity \cite{loop} or by the
unavoidable interaction with gravitons \cite{gravitons}. It is
therefore of interest to study the consequences induced by such
modifications into the dynamics of the scale factor of the
Universe. This is the main purpose of the present work. We will
consider  a quantum scalar field with a modified dispersion
relation on a Robertson Walker background, and analyze the
Semiclassical Einstein Equations (SEE) in order to test whether
the physics at very high scales may leave an imprint at lower
scales or not.

The source of the  SEE is the mean value of the energy momentum
tensor of the quantum fields, $\langle T_{\mu\nu}\rangle$, which
is  formally a divergent quantity. For quantum fields with usual
dispersion relations, the covariant  renormalization procedure is
very well known \cite{birrell,wald,fulling}. For Robertson Walker
metrics in $n$ dimensions,  one can compute the energy momentum
tensor in the so called  adiabatic approximation
\cite{staro,parkerhu,bunch}, and define the renormalized one as
\be\langle T_{\mu\nu}\rangle_{ren} = \langle T_{\mu\nu}\rangle -
\langle T_{\mu\nu}\rangle^{(0)}...
 - \langle T_{\mu\nu}\rangle^{(2[n/2])}\ee where $[n/2]$ is the integer part of $n/2$.
$\langle T_{\mu\nu}\rangle^{(j)}$ denotes the terms of adiabatic
order j of $\langle T_{\mu\nu}\rangle$ (i.e., the terms containing
$j$ derivatives of the metric) and once regularized is
proportional to a geometric tensor. This procedure is equivalent
to a redefinition of the gravitational bare constants of the
theory. In order to give sense to the divergent quantities
appearing in the equation above it is necessary to use a
regularization method, a very useful one being dimensional
regularization. In $3+1$ dimensions divergences arise up to the
fourth adiabatic order, and thus it is necessary to include in the
gravitational part of the theory terms quadratic in the curvature.
In toy $1+1$-dimensional models, the fourth adiabatic order is
finite, and the renormalization only involves the subtraction of
the zeroth and second adiabatic orders.

The SEE have also been considered for quantum fields with modified
dispersion relations \cite{lemoine,transpbr}. However, these previous
works  do not include a proper treatment of the divergences, and
the infinities are removed by simply neglecting the zero point
energy of each Fourier mode of the quantum fields. This approach
is not equivalent to a redefinition of the bare constants of the
theory, and therefore is not fully justified.

The extension of the adiabatic renormalization for generalized
dispersion relations was first considered by us in Ref.\cite{Nos},
where we computed the mean value of the energy momentum tensor up
to the second adiabatic order. Power counting arguments suggest
that, for dispersion relations containing $k^4$ or higher powers
of the momentum, it would be enough to subtract up to the second
adiabatic order in $3+1$ dimensions and up to the zeroth adiabatic
order in $1+1$ dimensions. However, a more careful analysis in
$1+1$ dimensions showed that this is not  the case \cite{NosProc}:
indeed, the quantities to be subtracted must be expressed in terms
of geometric tensors in $n$ dimensions, and the renormalization
should be performed before the limit $n\rightarrow 2$ is taken. In
$1+1$ dimensions $G_{\mu\nu}$ is proportional to $n-2$, since it
results from the variation of the would be Gauss-Bonnet invariant
in $n=2$. Thus, the mean value of the energy momentum tensor is
written as $\langle T_{\mu\nu}\rangle^{(2)} = c_n G_{\mu\nu}$
where the constant $c_n$ contains a pole at $n=2$. Therefore, the
second adiabatic order should also be subtracted in $1+1$,
although when explicitly computed it is a finite quantity. In this
way one obtains the correct trace anomaly and the results are
continuous in the limit $M_c\rightarrow\infty$ \cite{NosProc}. In
this paper we will show that the same situation arises in $3+1$
dimensions with the fourth adiabatic order. The tensor that
results from the variation of the would be Gauss-Bonnet invariant
at $n=4$ is proportional to $n-4$, and therefore $\langle
T_{\mu\nu}\rangle^{(4)}$, although finite, contains a pole when
expressed in terms of geometric tensors in $n$ dimensions. The
conclusion will be that the subtraction must also include the
fourth adiabatic order, whatever the dispersion relation.

As the calculations of the fourth adiabatic order are technically
rather involved, for the benefit of the reader we would like to
describe here with some detail the organization of the paper, our
main conclusions, and the relation with previous works on the
subject. In Section 2 we present the Lagrangian and energy
momentum tensor of the scalar field with modified dispersion
relations, as well as expressions for the geometric tensors in
$n-$dimensional Robertson Walker spacetimes that will be used
along the rest of the paper. Section 3 describes the WKB mode
functions for the scalar field, up to the fourth adiabatic order,
and for generic dispersion relations. For the usual dispersion
relation, the fourth adiabatic WKB mode functions have been
computed previously in Ref.\cite{bunch}, while for generic
dispersion relations, they have been computed up to the second
adiabatic order in Ref.\cite{Nos}.
 In Section 4 we use the WKB mode functions to construct the regularized
$\langle T_{\mu\nu}\rangle^{(j)}$, with $j=0,2,4$ in $n$
dimensions. These results generalize our previous ones for
$j=0,2$, and allow us to discuss one of the main points of this
paper, which is the necessity to subtract up to the fourth
adiabatic order in $n=4$. The calculations are relatively
straightforward but require a lot of algebra, so we relegate the
details to the Appendix A. In Section 5, and as a warm up, we
analyze the renormalization of the stress tensor in $n=2$. We
compute explicitly the trace of the energy-momentum tensor in de Sitter spacetime, and
show that it reproduces the usual value as $M_c\rightarrow\infty$.
Section 6 deals with the renormalization of $\langle
T_{\mu\nu}\rangle$ in $n=4$. Once more, as an example, the trace
of the energy-momentum tensor is computed in de Sitter space for a massless
conformally coupled quantum field, and the usual trace anomaly is
recovered in the limit $M_c\rightarrow\infty$.

In Section 7 we present a concrete application of the formalism
developed in previous sections. We first show that in de Sitter
space, for any dispersion relation, there exists a one complex
family of quantum states for which the unrenormalized energy
momentum tensor is proportional to $g_{\mu\nu}$. One member of the
family corresponds to a renormalizable state, and for this
particular state it is possible to obtain selfconsistent de Sitter
solutions, as explicitly shown for massless conformally coupled fields. We find that,
for values of the cosmological constant smaller than a critical
value, the SEE admit two different de Sitter solutions, a
perturbative one, close to the classical solution, and a
nonperturbative one, with a very high curvature.

In several previous works \cite{eff, Holman}, it has been argued
that the trans-Planckian physics could be taken into account by
considering an effective field theory with usual dispersion
relations,  in which the effects of the new physics is encoded in
the state of the field modes when they leave the
sub-Planckian regime. In Section 8 we discuss the possibility of
simulate the  modified dispersion relations by an effective
initial quantum state at the level of the SEE, i.e. to discuss the
backreaction effects.

Throughout the paper we set $c=1$ and adopt the sign convention
denoted (+++) by Misner, Thorne, and Wheeler \cite{MTW}.

\section{The model}

We consider a free quantum scalar field $\phi$ with modified dispersion relation
propagating in a curved  space-time with a classical spatially flat FRW metric given by \be
ds^2=g_{\mu\nu}dx^{\mu}dx^{\nu}\equiv
-(u_{\mu}dx^{\mu})^2+\perp_{\mu\nu}dx^{\mu}dx^{\nu}=C(\eta)[-d\eta^2+\delta_{i
j}dx^i dx^j]\label{metric}\ee where $\mu,\nu= 0,1...n-1$ (with
$n$  the space-time dimension), $C^{1/2}(\eta)$ is the scale factor given as
a function of the conformal time $\eta$, the vector field $u_{\mu}\equiv
C^{1/2}(\eta)\delta^{\eta}_{\mu}$, and $\perp_{\mu\nu}\equiv g_{\mu\nu}+ u_{\mu} u_{\nu}$ coincides with the
spatial metric as defined by an observer comoving with $u_{\mu}$.

The classical action for the scalar field  can be written as \cite{lemoine}: \be
S_{\phi}=\int d^n x \sqrt{-g}
(\mathcal{L}_{\phi}+\mathcal{L}_{cor}+\mathcal{L}_{u}),\ee where $g=det(g_{\mu\nu})$, $\mathcal{L}_{\phi}$ is the
standard Lagrangian of a free
 scalar field
\be \mathcal{L}_{\phi}=-\frac{1}{2}\lb g^{\mu
\nu}\partial_{\mu}\phi\partial_{\nu}\phi+(m^2+\xi R)\phi^2\rb,\ee
with $R$ the Ricci scalar, $\mathcal{L}_{cor}$ is the corrective lagrangian that gives rise
to a generalized dispersion relation \be
\mathcal{L}_{cor}=-\sum_{s,p\leq n} b_{sp}
(\mathcal{D}^{2s}\phi)(\mathcal{D}^{2p}\phi),\ee with
$\mathcal{D}^{2}\phi\equiv\perp_{\mu}^{\lambda}\nabla_{\lambda}\perp_{\gamma}^{\mu}\nabla^{\gamma}\phi$
(where
$\nabla_{\mu}$ is the covariant derivative corresponding to the metric $g_{\mu\nu}$ and $\perp_{\mu}^{\lambda}\equiv g^{\lambda\nu}\perp_{\mu\nu}$), and
$\mathcal{L}_{u}$ describes the dynamics of the additional degree
of freedom $u^{\mu}$ whose explicit expression is not necessary for
our present purposes.

The generalized dispersion relation takes the form \be \om^2_k=
k^2+C(\eta)\lb
m^2+2\sum_{s,p}(-1)^{s+p}\,b_{sp}\,\lp\frac{k}{C^{1/2}(\eta)}\rp^{2(s+p)}\rb,
\label{dis} \ee where  $b_{sp}$ are arbitrary coefficients, with
$p\leq s$.

The Fourier modes $\chi_k$ corresponding to the scaled field
$\chi=C^{(n-2)/4}\phi$ satisfy \be
\chi_k''+\lb(\xi-\xi_n)RC+\om_k^2\rb\chi_k=0, \label{PXXP}\ee with
the usual normalization condition \be \ck
{\ck'}^*-\ck'\ck^*=i.\label{nor} \ee Here primes stand for
derivatives with respect to the conformal time $\eta$, and in the conformal coupling case we have
$\xi=\xi_n\equiv(n-2)/(4n-4)$, while $\xi=0$ corresponds to
minimal coupling.

On the other hand, the gravitational action is given by
 \be
S_{G}=\frac{1}{16\pi G_N }\int d^n x \sqrt{-g}
(R-2\Lambda)-\frac{1}{2}\int d^n x \sqrt{-g}(\alpha R^2+\beta
R_{\mu\nu}R^{\mu\nu}+\gamma
R_{\mu\nu\rho\sigma}R^{\mu\nu\rho\sigma}),\label{Sg}\ee where
$R_{\mu\nu\rho\sigma}$ is the curvature tensor,
$R_{\mu\nu}=R^{\rho}_{\mu\rho\nu}$, and $\Lambda$, $G_N$,
$\alpha$, $\beta$, and $\gamma$ are bare parameters which are to
be appropriately chosen to cancel the corresponding divergences in
the $\langle T_{\mu\nu}\rangle$ derived from $S_{\phi}$. It is
well known that in $n=4$ dimensions, the quadratic terms in
Eq.(\ref{Sg}) are necessary for renormalizing the effective theory
in the case of scalar fields with the usual dispersion relation.
Moreover, as we will show, they are also necessary for any of the
generalized dispersion relations of the type given by
Eq.(\ref{dis}).

In the frame defined by the vector field
$u_{\mu}=C^{1/2}(\eta)\delta^{\eta}_{\mu}$, we can write the SEE
as \be \frac{1}{8 \pi G_N}(G_{\mu\nu}+\Lambda g_{\mu\nu})+\alpha
H_{\mu\nu}^{(1)}+\beta H_{\mu\nu}^{(2)}+\gamma H_{\mu\nu}=\langle
T_{\mu\nu}\rangle , \label{SEE}\ee where $\langle
T_{\mu\nu}\rangle$ is the expectation value of the energy momentum
tensor of the scalar field which in this frame satisfies $\langle
T^{\mu\nu}\rangle_{;\mu}=0$, $G_{\mu\nu}=R_{\mu\nu}-g_{\mu\nu}R/2$
is the Einstein tensor, and
\begin{subequations}
\begin{align}H_{\mu\nu}^{(1)}&=2 R_{;\mu\nu}-2 g_{\mu\nu}\Box R+\frac{1}{2} g_{\mu\nu} R^2-2 R R_{\mu\nu},\\
H_{\mu\nu}^{(2)}&=R_{;\mu\nu}-\frac{1}{2}  g_{\mu\nu}\Box R-\Box R_{\mu\nu} +\frac{1}{2} g_{\mu\nu} R_{\rho\sigma}R^{\rho\sigma}-2 R^{\rho\sigma} R_{\rho\mu\sigma\nu},\\
H_{\mu\nu}&=\frac{1}{2} g_{\mu\nu} R_{\rho\delta\sigma\gamma}
R^{\rho\delta\sigma\gamma}-2  R_{\mu\rho\delta\sigma} R_{\nu}{}
^{\rho\delta\sigma} -4 \Box R_{\mu\nu} + 2 R_{;\mu\nu}+4
R_{\mu\sigma}R^{\sigma}{}_{\nu}-4 R^{\rho\sigma}
R_{\rho\mu\sigma\nu}.
\end{align}\label{Haches}
\end{subequations}
The particular combination \be H_{\mu\nu}+H_{\mu\nu}^{(1)}-4
H_{\mu\nu}^{(2)}\equiv H_{\mu\nu}^{(3)}(n-4),\label{defH3}\ee
which comes from the variation of the would be Gauss-Bonnet
topological invariant at $n=4$, will be very important for our
discussion about the renormalizability of the theory in four
dimensions.

For the conformally flat metric that we are considering, the covariantly conserved tensors in Eq.(\ref{Haches}) are not independent,
\be 2 H_{\mu\nu}^{(1)}+(n-1)[(n-2) H_{\mu\nu}-4 H_{\mu\nu}^{(2)}]=0.\ee
 Therefore, it is enough to work with $H_{\mu\nu}^{(1)}$ and $H_{\mu\nu}^{(3)}$, whose non trivial components are given by
\begin{subequations}
\begin{align}
H_{\eta\eta}^{(1)}&=-\frac{(n-1)^2}{C}\left[\mathcal{H}\mathcal{H}''+\frac{(n-4)}{2} \mathcal{H}^2 \mathcal{H}'-\frac{\mathcal{H}'^2}{2}+\frac{(n-10)(n-2)}{32} \mathcal{H}^4\right],\label{Hetaeta1}\\
H_{11}^{(1)}&=\frac{2(n-1)}{C}\left[\mathcal{H}'''+\mathcal{H}'^2\left(\frac{1}{4}+\frac{3}{4}(n-4)\right)+\mathcal{H}\mathcal{H}''\left(-\frac{1}{2}+(n-4)\right)\right.\\
\nonumber & \left. +\mathcal{H}'\mathcal{H}^2\left(\frac{3}{8}(n-4)(n-6)-\frac{3}{2}\right)+\mathcal{H}^4\left(\frac{3}{16}+\frac{(n-4)}{64}(n^2-13 n+28)\right)\right],\\
H_{\eta\eta}^{(3)}&=-\frac{(n-1)(n-2)(n-3)}{32 C}  \mathcal{H}^4,\label{Hetaeta3}\\
H_{11}^{(3)}&=\frac{(n-2)(n-3)}{4 C}\left[\mathcal{H}'\mathcal{H}^2+\frac{(n-5)}{8}\mathcal{H}^4\right],
\end{align}\label{HachesRW}
\end{subequations} with $H_{11}^{(1,3)}=H_{22}^{(1,3)}=...=H_{(n-1)(n-1)}^{(1,3)}$, and $\mathcal{H}\equiv C'/C$.

The nontrivial components of the Einstein tensor are
\begin{subequations}
\begin{align}
G_{\eta\eta}=&\ \frac{(n-1)}{4}\frac{(n-2)}{2}\mathcal{H}^2, \\
G_{11}=G_{22}=...=G_{(n-1)(n-1)}
=&-\frac{(n-2)}{2}\lb\mathcal{H'}+\frac{(n-3)}{4}\mathcal{H}^2\rb,
\end{align}\label{EinsteinTensorRW}
\end{subequations}
and the Ricci scalar takes the form
\begin{equation}\label{RicciRW}
R=\frac{(n-1)}{C}\left[\mathcal{H}'+\frac{(n-2)}{4}\mathcal{H}^2\right].
\end{equation}

The expectation value of the energy momentum tensor of the scalar field can be written as \cite{signo}:
\begin{eqnarray}
\nonumber\langle T_{\eta\eta}\rangle &=&  \sqrt{C}\int
\frac{d^{n-1}k\,\mu^{\bar{n}-n}}{(2\pi\sqrt{C})^{(n-1)}} \left\{
\frac{C^{(n-2)/2}}{2}\left|\lp \frac{\chi_k}{C^{(n-2)/4}}\rp'
\right|^2 +\frac{\om_k^2}{2}\,|\chi_k|^2+\xi
G_{\eta\eta}|\chi_k|^2\right.
\\ &+&\left.\xi\frac{(n-1)}{2}
\lb\frac{C'}{C}(\chi_k'\chi_k^*+\chi_k{\chi_k'}^{*})-\frac{{C'}^2}{C^2}\frac{(n-2)}{2}|\chi_k|^2
\rb \right\},\label{RHOO}\\
\nonumber \langle T_{11}\rangle  &=&  \sqrt{C}\int
\frac{d^{n-1}k\,\mu^{\bar{n}-n}}{(2\pi\sqrt{C})^{(n-1)}}
\left\{\lp\frac{1}{2}-2\xi\rp C^{(n-2)/2}\left|\lp
\frac{\chi_k}{C^{(n-2)/4}}\rp' \right|^2+\xi
G_{11}|\chi_k|^2\right.\\\nonumber
 &+&\lb \lp\frac{k^2}{n-1}\rp\frac{d\om_k^2}{
dk^2}-\frac{\om_k^2}{2}\rb|\chi_k|^2-\xi(\chi_k''\chi_k^*+\chi_k{\chi_k''}^{*})+\xi\frac{
C'}{2C}(\chi_k'\chi_k^*+\chi_k{\chi_k'}^{*})\\
&+&\left.\xi\frac{(n-2)}{2}\lp\frac{C''}{C}-\frac{(8-n)}{4}\frac{{C'}^2}{C^2}\rp|\chi_k|^2\right\},\label{PPP}
\end{eqnarray} with $T_{11}=T_{22}=...=T_{(n-1)(n-1)}$. Here $\bar{n}$ is the physical space-time dimension, and $\mu$ is an arbitrary parameter with mass dimension introduced to ensure that $\chi$ has the correct dimensionality.

\section{The WKB expansion}

In order to compute the WKB expansion, we begin by expressing
the field mode $\chi_k$ in the well-known form \be
\ck= \frac{1}{\sqrt{ 2 W_k}}\exp\lp -i\int^\eta
W_k(\tilde\eta)d\tilde\eta\rp .\label{chi} \ee
Then, the expectation values of the stress tensor can be written as
\begin{eqnarray}
\nonumber\langle T_{\eta\eta}\rangle &=& \Omega_{n-1} \frac{\sqrt{C}}{
2}\int \frac{dk\,k^{n-2}\,\mu^{\bar{n}-n}}{(2\pi\sqrt{C})^{n-1}}\left\{\frac{[(W_{k}^2)']^2}{32 W_{k}^5}+\frac{W_{k}}{ 2}+\frac{\om_k^2}{ 2W_{k}}+\frac{(n-2)}{2}\lb\frac{ {C'}^2(n-2)}{16 W_{k}C^2}+\frac{ {C'}(W_{k}^2 )'}{8CW_{k}^3}\rb\right.\\
&+&\left.\xi
\frac{G_{\eta\eta}}{W_{k}}-\xi\frac{(n-1)}{2}\lb\frac{{C'}^2}{
C^2}\frac{(n-2)}{2 W_{k}}+\frac{C'}{C}\frac{(W_{k}^2)'}{2 W_{k}^3}\rb\right\},
\label{rhoad}
\end{eqnarray}
\begin{eqnarray}
\langle T_{11}\rangle &=& \Omega_{n-1} \frac{\sqrt{C}}{
2}\int \frac{dk\,k^{n-2}\,\mu^{\bar{n}-n}}{(2\pi\sqrt{C})^{n-1}}\left\{ \frac{[(W_{k}^2)']^2}{ 32 W_{k}^5}+\frac{W_{k}}{2}-\frac{\om_k^2}{2W_{k}}+\frac{(n-2)}{2}\lb\frac{ {C'}^2(n-2)}{ 16 W_{k} C^2}+\frac{ {C'}(W_{k}^2 )'}{ 8CW_{k}^3}\rb \right.\label{pe}\nonumber\\
&+&\frac{k^2}{(n-1) W_{k}}\frac{d\om^2}{
dk^2}+\xi\frac{G_{11}}{W_{k}}+\xi\lb\frac{(W_{k}^2)''}{2
W_{k}^3}-\frac{3}{4}\frac{[(W_{k}^2)']^2}{W_{k}^5}-\frac{(n-1)}{4}\frac{C'(W_{k}^2)'}{C
W_{k}^3} \rb\nonumber\\
&+&\left.\frac{(n-2)}{2}\frac{\xi}{
W_{k}}\lb\frac{C''}{C}-\frac{3}{2}\frac{{C'}^2}{C^2}\rb\right\},\label{pad}
\end{eqnarray}
where we have defined  the factor $\Omega_{n-1}\equiv
2\pi^{(n-1)/2}/\Gamma[(n-1)/2]$ coming from the angular
integration.

Substitution of Eq.(\ref{chi})
into Eq.(\ref{PXXP}) yields
\begin{equation}
W_k^2  =\Omega_k^2-\frac{1}{2}\lp\frac{W_k''}{W_k}-\frac{3}{2}\frac{{W'_k}^2}
{W_k^2}\rp,\label{W}
\end{equation}
where $\Omega_k^2 =  \om_k^2+\lp\xi-\xi_n\rp
CR$. This non-linear differential equation can be solved iteratively for $W_k$ by assuming that it is a slowly varying function of
$\eta$. In this adiabatic or WKB approximation, the adiabatic order
of a term is given by the number of time derivatives.

In what follows we will denote by ${}^{(j)}W_k^2$ the terms in $W_k^2$ of adiabatic order j.
We obtain straightforwardly  ${}^{(2)}W_k^2$
replacing $W_k$ by $\Omega_k$ on the right hand side of Eq.(\ref{W}),
\begin{eqnarray}
{}^{(2)}W_k^2  &= &
\lp \xi-\xi_n\rp(n-1)\lp\frac{C''}{C}+\frac{(n-6)}{4}\frac{{C'}^2}{C^2}\rp \nonumber\\
& -& \   \frac{1}{4}\frac{C''}{C}\lp 1-\frac{k^2}{\om_k^2}\frac{d\om_k^2}{d k^2}\rp-\frac{1}{4}\frac{{C'}^2}{C^2}\frac{k^4}{\om_k^2}\frac{d^2\om_k^2}{{d(k^2)}^2}\nonumber\\
& +&  \frac{5}{16}\frac{{C'}^2}{C^2}\lp1-\frac{k^2}{\om_k^2}\frac{d\om_k^2}{d k^2}
\rp^2\nonumber\\
&=& \frac{\mathcal{H}^2}{16}[f^2-4 \dot{f}+4(\xi-\xi_n)(n-1)(n-2)]+\frac{\mathcal{H}'}{4}[f+4(\xi-\xi_n)(n-1)],\label{W2ad2}
\end{eqnarray}where we have used the fact that $\tilde{\omega}_k^2\equiv\omega_k^2/C$ depends on time only through the variable
$x\equiv k^2/C$ to rewrite the right hand side in terms of the function
\be f\equiv \frac{d\ln\tilde{\omega}_k^2}{d\ln x}-1,\label{f}
\ee and a dot means a derivative with respect to $\ln x$.

For $\bar{n}=4$ dimensions we will also need the fourth adiabatic order, therefore, we perform one more iteration in Eq.(\ref{W}). Discarding the higher order terms we arrive at
\begin{equation}
{}^{(4)}W_k^2= \frac{1}{8\omega_k^2} \left\{ {}^{(2)}W_k^2[\mathcal{H}^2(2 \dot{f}-3f^2)-2\mathcal{H}' f]-5{}^{(2)}{W_k^2}'\mathcal{H}f-2{}^{(2)}{W_k^2}''\right\},\label{W2ad4}
\end{equation}where we have used that $(\omega^2_k)'/\omega^2_k=-\mathcal{H}f$ and $(\omega^2_k)''/\omega^2_k=\mathcal{H}^2(f^2+\dot{f})-\mathcal{H}'f$.

In what follows it will be relevant to know the dependence with $k$ of the different
adiabatic orders. From Eqs. (\ref{W}), (\ref{W2ad2}) and (\ref{W2ad4}), with the use of an inductive argument, it can be shown that the $2j-$adiabatic order scales as $\om_k^{2-2j}$.

\section{Regularized adiabatic stress tensor}

In this Section we present the regularized adiabatic stress
tensor, up to the fourth order, in $n$ dimensions. Our goal will
be to show that the different adiabatic orders are proportional to
the conserved tensors $g_{\mu\nu}, G_{\mu\nu}$ and
$H_{\mu\nu}^{(i)}$ that appear into the SEE. Therefore its
subtraction will be equivalent to a redefinition of the bare
gravitational constants of the theory.

In order to find the tensorial structure of the different
adiabatic orders, we will perform several integrations by parts in
the integrals appearing in the WKB expansion of the stress tensor,
and use the fact that in dimensional regularization the integral
of a total derivative vanishes \cite{Collins}. We will sketch here
the calculations, leaving the details to the Appendix A.

The zeroth and second adiabatic orders of the expectation value of the stress tensor
can be evaluated from  Eqs. (\ref{rhoad}), (\ref{pad}) and (\ref{W2ad2}). They
have been computed in Ref.\cite{Nos}, and are given by  \cite{error}:
\begin{subequations}
\begin{align}
\langle T_{\eta\eta}\rangle^{(0)} &= \frac{C}{
4}\frac{\Omega_{n-1}\,\mu^{\bar n-n}}{(2\pi)^{n-1}} \int_0^{\infty} dx \,x^\frac{(n-3)}{2}\tilde{\omega}_k,\\
\langle T_{11}\rangle^{(0)} &= \frac{C}{
4}\frac{\Omega_{n-1}\,\mu^{\bar n-n}}{(2\pi)^{n-1}} \int_0^{\infty} dx \,x^\frac{(n-3)}{2}\frac{(f+1)}{n-1}\tilde{\omega}_k;
\end{align}
\end{subequations}
\begin{subequations}
\begin{align}
\langle T_{\eta\eta}\rangle^{(2)}& = \frac{\Omega_{n-1}\,\mu^{\bar n-n}}{4 (2\pi)^{n-1}} \int_0^{\infty} dx\, \frac{x^\frac{(n-3)}{2}}{\tilde{\omega}_k}\left\{\frac{\mathcal{H}^2}{32}(f-n+2)^2+ \xi\frac{\mathcal{H}^2}{4}(n-1) (f-n+2) +\xi G_{\eta\eta}\right\},\\
\langle T_{11}\rangle^{(2)}& = \frac{\Omega_{n-1}\,\mu^{\bar n-n}}{4(2\pi)^{n-1}} \int_0^{\infty} dx\, \frac{x^\frac{(n-3)}{2}}{\tilde{\omega}_k}\left\{\frac{(f-n+2)^2}{32(n-1)}[\mathcal{H}^2(n-1)-4\mathcal{H}']+\frac{\mathcal{H}^2 (f-n+2)}{32 (n-1)}\right.\\\nonumber
\times &\left.[4\dot{f}-f^2+(n-2)^2]+\frac{\xi\mathcal{H}^2}{8}[4\dot{f}-2f^2+n f+(n-2)(n-4)]-\xi \mathcal{H}' (f-n+2)+\xi G_{11}\right\}.
\end{align}
\end{subequations}
Performing several integrations by parts and discarding surface terms, one obtains \cite{Nos}:
 \be \langle
T_{\mu\nu}\rangle^{(0)} = - \frac{g_{\mu\nu}}{
4}\frac{\Omega_{n-1}\,\mu^{\bar n-n}}{(2\pi)^{n-1}} I_0\, ,\label{Tmunuad0}\ee
\be
\langle T_{\mu\nu}\rangle^{(2)}  =
G_{\mu\nu}\frac{\Omega_{n-1}\,\mu^{\bar n
-n}}{4\,(2\pi)^{n-1}}\left\{
  \frac{I_2}{6(n-1)(n-2)}+(\xi - \frac{1}{6}) I_1\right\},\label{Tmunuad2}
\ee where we have denoted by
 $I_i$ (i=0,1,2) the integrals given in Table \ref{tabla}. Eqs.
 (\ref{Tmunuad0}) and (\ref{Tmunuad2}) show explicitly that the
 zeroth and second adiabatic orders can be absorbed into a
 redefinition of the bare cosmological and Newton constants
 respectively.

\begin{table}[h]
\begin{tabular}{|c|c|}
\hline
& \\
$I_0 =\int_0^\infty dx\,x^{\frac{(n-3)}{2}}{\tilde{\om}_k}$ &   $I_5=\int_{0}^{\infty} dx \frac{x^{\frac{(n+3)}{2}}}{\tilde{\omega}_k^5} \frac{d^3\tilde{\omega}_k^2}{{dx}^3}$        \\
& \\
\hline
& \\
$I_1=\int_0^\infty dx\,\frac{x^{\frac{(n-3)}{2}}}{\tilde{\om}_k}$& $I_6=\int_{0}^{\infty} dx \frac{x^{\frac{(n+5)}{2}}}{\tilde{\omega}_k^5}\frac{d^4\tilde{\omega}_k^2}{{dx}^4}$\\
& \\
\hline
& \\
$I_2 = \int_0^\infty dx\,\frac{x^{\frac{(n+1)}{2}}}{\tilde{\om}_k^3}\frac{d^2\tilde{\om}_k^2}{{dx}^2}$ & $I_7=\int_{0}^{\infty} dx \frac{x^{\frac{(n+5)}{2}}}{\tilde{\omega}_k^7}\left(\frac{d^2\tilde{\omega}_k^2}{{dx}^2}\right)^2$\\
& \\
\hline
& \\
$I_3=\int_{0}^{\infty} dx \frac{x^{\frac{(n-3)}{2}}}{\tilde{\omega}_k^3}$& $I_8=\int_{0}^{\infty} dx \frac{x^{\frac{(n+7)}{2}}}{\tilde{\omega}_k^5} \frac{d^5\tilde{\omega}_k^2}{{dx}^5}$  \\
& \\
\hline
& \\
$I_4=\int_{0}^{\infty} dx \frac{x^{\frac{(n+1)}{2}}}{\tilde{\omega}_k^5} \frac{d^2\tilde{\omega}_k^2}{{dx}^2}$  &
$I_9=\int_{0}^{\infty} dx \frac{x^{\frac{(n+7)}{2}}}{\tilde{\omega}_k^7}\frac{d^3\tilde{\omega}_k^2}{{dx}^3}\frac{d^2\tilde{\omega}_k^2}{{dx}^2}$ \\
& \\
\hline
\end{tabular}
 \caption{Explicit expressions for $I_{i}$. To obtain these integrals we have made the change of variables  $x\equiv k^2/C$ and we have defined $\tilde{\om}_k=\om_k/\sqrt{C}$.}\label{tabla}
\end{table}

The fourth adiabatic order can be computed by following the same
procedure. Starting from Eqs. (\ref{rhoad}) and (\ref{pad}) for
$\langle T_{\eta\eta}\rangle$ and $\langle T_{11} \rangle$, we use
the adiabatic expansions given in Eqs. (\ref{W2ad2}) and
(\ref{W2ad4}) to arrive at the following expressions for the
fourth adiabatic order of these expectation values:
\begin{subequations}\label{tcoeff}
\begin{align}
\langle T_{\eta\eta}\rangle^{(4)} &=\frac{\Omega_{n-1}\mu^{\bar{n}-n}}{4 C (2\pi)^{(n-1)}}[ \alpha_1\mathcal{H}'\mathcal{H}^2+\alpha_2\mathcal{H}''\mathcal{H}+\alpha_3\mathcal{H}^4+\alpha_4{\mathcal{H}'}^2],\\
\langle T_{11}\rangle^{(4)} &=\frac{\Omega_{n-1}\mu^{\bar{n}-n}}{4 C (2\pi)^{(n-1)}}[ \beta_1\mathcal{H}'\mathcal{H}^2+\beta_2\mathcal{H}''\mathcal{H}+\beta_3\mathcal{H}^4+\beta_4{\mathcal{H}'}^2+\beta_5\mathcal{H}'''].
\end{align}
\end{subequations}
The coefficients $\alpha_i$ and $\beta_i$ are given in terms of integrals that involve
${\tilde\omega}_k$ and its derivatives. As anticipated, one can find relations
between them in order to show explicitly the geometric structure of the adiabatic stress tensor.
To carry out this procedure, it is convenient to express all these coefficients in terms of the integrals $I_i$ defined in Table \ref{tabla}. For example, let us consider the coefficient $\alpha_2$,
\begin{eqnarray}\label{alpha2}
\alpha_2&=&-\frac{1}{64}\int_{0}^{+\infty}\frac{dx\, x^{\frac{(n-3)}{2}}}{\tilde{\omega}_k^3}[4(n-1)(\xi-\xi_n)+f]^2\nonumber\\
&=&-\frac{1}{4}(n-1)^2(\xi-\xi_n)^2 I_3-\frac{1}{8}(n-1)(\xi-\xi_n)\int_{0}^{+\infty}\frac{dx\, x^{\frac{(n-3)}{2}}}{\tilde{\omega}_k^3}f-\frac{1}{64}\int_{0}^{+\infty}\frac{dx\, x^{\frac{(n-3)}{2}}}{\tilde{\omega}_k^3}f^2.
\end{eqnarray} Taking into account the definition of $f$ (Eq.(\ref{f})) we have,

\begin{subequations}\label{J12000}
\begin{align}
\int_{0}^{+\infty}\frac{dx\, x^{\frac{(n-3)}{2}}}{\tilde{\omega}_k^3}f&=-\frac{2}{3}\int_{0}^{\infty}x^{\frac{(n-1)}{2}}\frac{d\tilde{\omega}_k^{-3}}{dx\,\,}-I_3=\frac{(n-4)}{3}I_3,\\
\int_{0}^{+\infty}\frac{dx\, x^{\frac{(n-3)}{2}}}{\tilde{\omega}_k^3}f^2&=-\frac{2}{5}\int_{0}^{\infty}x^{\frac{(n+1)}{2}}\frac{d\tilde{\omega}_k^{-5}}{dx\,\,}\frac{d\tilde{\omega}_k^{2}}{dx\,}+\frac{4}{3}\int_{0}^{\infty}x^{\frac{(n-1)}{2}}\frac{d\tilde{\omega}_k^{-3}}{dx\,\,}+I_3\nonumber\\
&=\frac{1}{15}(n-4)(n-6)I_3+\frac{2}{5}I_4.
\end{align}
\end{subequations}
and therefore
\begin{equation}
\alpha_2= -\frac{(n-1)^2}{4}\left(\xi-\frac{1}{6}\right)^2 I_3+\frac{(n-4)(n-1) }{1440} I_3-\frac{I_4}{160}.
\end{equation}

Using a similar procedure, outlined in Appendix A,  it can be
shown that the coefficients are related in such a way that the
regularized expression for $\langle T_{\mu\nu}\rangle^{(4)}$ takes
the form \be \langle T_{\mu\nu}\rangle^{(4)}= B_1
H_{\mu\nu}^{(1)}+B_3 H_{\mu\nu}^{(3)},\label{Tmunuad4} \ee where
$H_{\mu\nu}^{(1)}$ and $H_{\mu\nu}^{(3)}$ are the tensors given in
Eq.(\ref{HachesRW}), and
\begin{subequations}
\begin{align}
B_1&= \frac{\Omega_{n-1}\mu^{\bar{n}-n}}{4 (2\pi)^{(n-1)}}\left\{\frac{I_3}{4}\left[
\left(\xi-\frac{1}{6}\right)^2+\frac{(n-4)}{360 (n-1)}\right]+\frac{I_4}{160(n-1)^2}\right\}\label{coefB1}\\\label{coefB3}
B_3&= \frac{\Omega_{n-1}\mu^{\bar{n}-n}}{4 (2\pi)^{(n-1)}}\left\{\frac{I_3(n-6)}{1440(n-3)}+\frac{I_4}{(n-4)(n-3)}\left[
\frac{(n+2)}{4(n-2)}(\xi-\xi_n)^2+\frac{9(n+26)}{1440(n-1)}\right]\right.\\\nonumber
&+\left.\frac{3}{1440}\frac{(32 I_5(n-5)+60 I_6-210 I_7)}{(n-1)(n-2)(n-3)(n-4)}\right\},
\end{align}\label{coeficientes}
\end{subequations}where all the integrals are shown in Table
\ref{tabla}. From Eq.(\ref{Tmunuad4}) we see that the fourth
adiabatic order of the energy momentum tensor can be absorbed into
redefinitions of the bare gravitational constants $\alpha$,
$\beta$, and $\gamma$ of the SEE (\ref{SEE}).

\section{Renormalization of the stress tensor in 1+1 dimensions}

Knowing the dependence with $k$ of the $2j-$adiabatic orders, one
can use power counting arguments to see which of the adiabatic
orders contain divergences. In $1+1$ dimensions, it is simple to
check that for any of the dispersion relations we are considering,
even for the usual one,  the zeroth adiabatic order is divergent,
while the higher orders are finite. This suggests that only a
redefinition of the cosmological constant would be required to
absorb the divergences and, then, one would define the
renormalized stress tensor as  $\langle
T_{\mu\nu}\rangle_{ren}=\langle T_{\mu\nu}\rangle-\langle
T_{\mu\nu}\rangle^{(0)}$. However, as will be explained below,
this naive argument is incorrect. For the usual dispersion
relation, it is known that to obtain the renormalized stress
tensor it is necessary to subtract not only the zeroth adiabatic
order, but also the second one \cite{birrell}.  As for the usual
dispersion relation $I_2=0$, from Eq.(\ref{Tmunuad2}) we obtain
\begin{eqnarray}
\langle T_{\mu\nu}\rangle^{(2)}&=&
G_{\mu\nu}\frac{\Omega_{n-1}\,\mu^{\bar n
-n}}{4\,(2\pi)^{n-1}}(\xi - \frac{1}{6})
I_1\label{orden2usual}\\\nonumber &=& -
\frac{G_{\mu\nu}}{2\pi}(\xi -
\frac{1}{6})\left[\frac{1}{n-2}+\ln\left(\frac{m}{2\mu}\right)+\mathcal{O}(n-2)\right].\end{eqnarray}
The point is that the Einstein tensor (which results from the
variation of the would be Gauss-Bonnet topological invariant at
n=2) vanishes as $n\to 2$: $G_{\mu\nu}\propto n-2$ (see
Eq.(\ref{EinsteinTensorRW})). Therefore, from
Eq.(\ref{orden2usual}) we see that the second adiabatic order of
the $n$-dimensional $\langle T_{\mu\nu}\rangle$ is finite in the
limit $n\to 2$. However, when  written in terms of $G_{\mu\nu}$ an
explicit pole at $n=2$ appears. As the pole must be absorbed into
the bare constant before taking the limit $n\to 2$, a redefinition
of the bare Newton constant is also required. In fact, the
contribution of the second adiabatic order  yields the well known
trace anomaly in the case of a massless, conformally coupled field
($\xi=0$) \cite{birrell}: \be\langle T^{\mu}_{\mu}
\rangle_{ren}=-\langle T^{\mu}_{\mu}\rangle^{(2)}=
\frac{R}{24\pi}.\label{usual}\ee

On the other hand, for a generalized dispersion relation such that
$\omega_k^2\sim k^{2r}$ with $r>1$, the integral $I_1$ is finite,
but the pole appears explicitly in Eq.(\ref{Tmunuad2}) multiplying
$I_2$ (which is also finite). Note that the subtraction of
$\langle T_{\mu\nu}\rangle^{(2)}$ can be consistently done
provided that it is proportional to $G_{\mu\nu}$ and hence the
pole can be absorbed into the bare Newton constant. Since the
renormalization prescription must be equivalent to a redefinition
of the bare constants in the effective Lagrangian of the theory,
we conclude that the second adiabatic order should also  be
subtracted in this case. Therefore, the renormalized stress tensor
is given by \cite{NosProc} \be\langle T_{\mu\nu}
\rangle_{ren}=\langle T_{\mu\nu}\rangle-\langle
T_{\mu\nu}\rangle^{(0)}-\langle
T_{\mu\nu}\rangle^{(2)}.\label{presc}\ee

As a consistency check of this renormalization prescription, let
us consider a massless field with $\xi=\xi_2=0$ in de Sitter
space-time, $C(\eta)=\alpha^2/\eta^2$, and a particular dispersion
relation of the form $\omega_k^2=k^2+2b_{11}k^4/C(\eta)$, with
$b_{11}>0$. We will compute the trace anomaly by taking the limit
in which the dispersion relation tends to the usual one
($b_{11}\to 0$). If we only subtract the zeroth adiabatic order,
we have
\begin{equation}
\langle T^{\mu}_{\mu}\rangle- \langle T^{\mu}_{\mu}\rangle^{(0)} =
-\frac{1}{\pi C} \int_0^\infty dk
\left(1-\frac{k^2}{\om_k^2}\frac{d\om_k^2}{dk^2}\right)\left(\om_k^2\vert
\chi_k\vert^2 -\frac{\om_k}{2}\right)
\end{equation}
The modes of the field satisfy (see Eq.(\ref{PXXP}))
\be \frac{\partial ^2\ck}{ \partial \eta^2}+\lp k^2+\frac{2b_{11}k^4 \eta^2}{\alpha^2}\rp\ck=0. \ee
 With the substitution $s=(2b_{11})^{1/4}\alpha^{-1/2}k\eta$ and introducing the constant
$\la=\alpha (2b_{11})^{-1/2}$, the equation becomes \be
\frac{\partial ^2\ck}{ \partial s^2}+\lp \la+ s^2\rp\ck=0. \ee The
particular solution of this equation that satisfies the
normalization condition (\ref{nor}) and tends to the adiabatic
mode of positive frequency for $|s| \to \infty$
 ($\eta\to - \infty$) is given by \cite{Nos}:
 \be \chi_k(s)=\frac{e^{-\lambda\pi/8}}{\sqrt{k}}
\left(\frac{\lambda}{2}\right)^{1/4} D_{-\left(\frac{1-i\lambda}{ 2}\right)}\left[(1-i)s\right],\label{sol} \ee
 where $D$ is the
parabolic function \cite{GR,Abram}, $s=k\eta/\sqrt{\lambda}$ and
$\lambda=\alpha/\sqrt{2 b_{11}}$. After changing variables and some
algebra we get
\begin{equation} \langle T^{\mu}_{\mu}\rangle- \langle T^{\mu}_{\mu}\rangle^{(0)} = \frac{R}{2\pi }\int_0^{\infty} ds
s^3\left\{f(\lambda,s)-\frac{\sqrt{\lambda}}{2\sqrt{\lambda+s^2}}\right\},\label{tm}
\end{equation} where $f(\lambda,s)\equiv k |\chi_k(s)|^2$ and $R=12\alpha^{-2}$.
A numerical evaluation of this integral gives, in the limit
$b_{11}\to 0$, \be \langle T^{\mu}_{\mu}\rangle-
\langle T^{\mu}_{\mu}\rangle^{(0)}\to -\frac{R}{24\pi}. \label{contribnum}\ee  As for the case of the usual
dispersion relation ($b_{11}=0$), the trace of the stress tensor has an
anomaly.  However, the numerical value does not coincide with the
usual one (it differs by a sign). Therefore, if we subtract only
the zeroth adiabatic order, there is a discontinuity in the
renormalized stress tensor as $b_{11}\to 0$. This
discontinuity disappears if we also subtract the second adiabatic order. Indeed, from Eq.(\ref{Tmunuad2}) we find, near $n=2$,
\begin{equation}
\langle T^{\mu}_{\mu}\rangle^{(2)}=-\frac{R}{48\pi}\left((2-n)I_1+I_2\right)\mu^{2-n}\label{t2}
\end{equation}
As $I_1$ is finite for non vanishing coefficient $b_{11}$, the
first term does not contribute to the trace in $n=2$. On the other
hand, $I_2$ is independent of $b_{11}$, and an explicit evaluation
gives \be \langle
T^{\mu}_{\mu}\rangle^{(2)}=-\frac{R}{12\pi}.\label{contribAd2} \ee
So, combining Eqs. (\ref{presc}), (\ref{contribnum}) and
(\ref{contribAd2}) we see that the usual trace anomaly
(\ref{usual}) is recovered in the limit  $b_{11}\to 0$, \be
\langle T^{\mu}_{\mu}\rangle_{ren}=\langle T^{\mu}_{\mu}\rangle-\langle T^{\mu}_{\mu}\rangle^{(0)}-\langle T^{\mu}_{\mu}\rangle^{(2)}\to \frac{R}{24\pi}.\ee

\begin{figure}[htp]
\includegraphics[width=8cm]{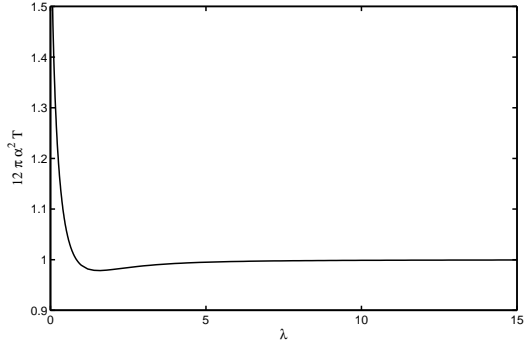}
\caption{The renormalized trace of the stress tensor $T$ (with the
prescription in Eq.(\ref{presc})) normalized to its anomalous
value as a function of $\lambda=\alpha/\sqrt{2 b_{11}}$. These
numerical results correspond to the massless scalar field with
$\xi=0$ and $\omega_k^2=k^2+2b_{11}k^4/C(\eta)$ propagating in a
two dimensional de Sitter background. }\label{traza2d}
\end{figure}

For the sake of completeness,  we compute the trace of the stress
tensor ($T$), renormalized according to the prescription in
Eq.(\ref{presc}), for all values of $b_{11}$ (we recall that the
limit $m\to 0$ has to be taken at the end of the calculations). In
Fig. \ref{traza2d} we have plotted the trace $T$ as a function of
$\lambda=\alpha/\sqrt{2 b_{11}}$.  In this figure we see that as
$\lambda$ increases ($b_{11}$ decreases) the trace approaches its
anomalous value.

\section{Renormalization of the stress tensor in 3+1 dimensions}

In this Section, we will show that in $3+1$ dimensions the fourth adiabatic order can be consistently subtracted, and that
this must be done for any of the dispersion relations of the form given in Eq.(\ref{dis}).

In $3+1$ dimensions, by power counting one can show that for
$\om_k^2\sim k^{2r}$, with $r\geq 4$, all contributions of second
or higher adiabatic orders are finite. The divergences come only
from the zeroth order terms contained in $\langle
T_{\mu\nu}\rangle$. In the cases $\om_k^2\sim k^6$ and
$\om_k^2\sim k^4$, though no fourth order divergences appear,
second order terms include divergent contributions, which suggests
that no terms quadratic in the curvature would be necessary in the
SEE. However, the situation in $3+1$ dimensions is similar to that
in $1+1$ \cite{f1}. By using the definition of $H_{\mu\nu}^{(3)}$
given in Eq.(\ref{defH3}), we can rewrite Eq.(\ref{Tmunuad4}) as
\be \langle T_{\mu\nu}\rangle^{(4)}= B_1
H_{\mu\nu}^{(1)}+\frac{B_3}{(n-4)} \left[
H_{\mu\nu}+H_{\mu\nu}^{(1)}-4 H_{\mu\nu}^{(2)}\right]. \ee As the
coefficient $B_3$ does not vanish in four dimensions, an explicit
pole at $n=4$ appears, which must be absorbed into the bare
constants of the effective gravitational action (\ref{Sg}).

If the fourth adiabatic order is not subtracted, there would be a
discontinuity in the limit in which the dispersion relation tends
to the usual one. This means that the renormalized stress tensor
would contain non vanishing trans-Planckian contributions, even
when $M_C\to \infty$. We illustrate this point by computing the
trace of the energy momentum tensor for a massless field with
conformal coupling $\xi=1/6$ and
$\omega_k^2=k^2+2b_{11}k^4/C(\eta)$ in de Sitter space-time,
$C(\eta)=\alpha^2/\eta^2$ (as we have done for the $1+1$
dimensional case). From Table \ref{tabla} one can see that for the
usual dispersion relation the only divergent integral appearing in
the fourth order is $I_3$  (see Eqs. (\ref{Tmunuad4}) and
(\ref{coeficientes})). So, near $n=4$, we have \be \langle
T_{\mu\nu}\rangle^{(4)}\to\frac{1}{2880\pi^2}\left[\frac{1}{6}{}^{(1)}H_{\mu\nu}+{}^{(3)}
H_{\mu\nu}\right], \ee
%,\,\,\,\,\left(\xi=\frac{1}{6},\,\omega^2_k=k^2+C m^2,\mathrm{ and }\,\,m\to 0\right)
which is a well known result \cite{birrell,bunch}. Therefore, the usual trace anomaly is given by \be
\langle T_{\mu}^{\mu}\rangle_{ren}=-\langle T_{\mu}^{\mu}\rangle^{(4)}=-\frac{R^2}{34560\pi^2}=-\frac{1}{240\pi^2\alpha^4}.
\ee
%,\,\,\,\,\,\left(\xi=\frac{1}{6},\,\omega^2_k=k^2+C m^2,\mathrm{ and }\,\,m\to 0\right)
On the other hand, for $b_{11}>0$ all integrals are finite and can
be explicitly computed. We find \be\label{orden4} \langle
T_{\mu}^{\mu}\rangle^{(4)}=\frac{1}{120\pi^2\alpha^4}, \ee which
is $-2$ times the usual trace anomaly. Note that the result is
independent of $b_{11}$.

\begin{figure}[htp]
\includegraphics[width=8cm]{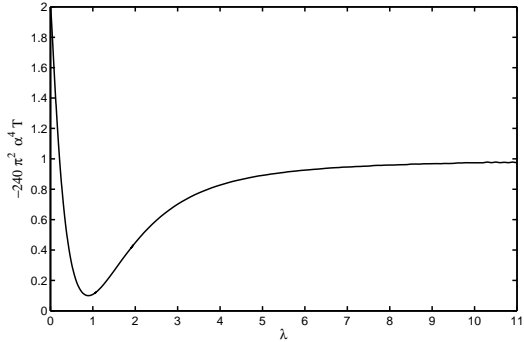}
\caption{The renormalized trace of the stress tensor normalized to
its anomalous value as a function of $\lambda=\alpha/\sqrt{2
b_{11}}$. The numerical result corresponds to the massless
scalar field with  $\xi=1/6$ and
$\omega_k^2=k^2+2b_{11}k^4/C(\eta)$ propagating in a four
dimensional de Sitter background. }\label{traza4d}
\end{figure}
The trace of the unrenormalized stress tensor with the subtraction of the zeroth and second adiabatic orders can be written in the form
\be\label{orden0y2}\langle T_{\mu}^{\mu}\rangle-\langle T_{\mu}^{\mu}\rangle^{(0)}-\langle T_{\mu}^{\mu}\rangle^{(2)}=\frac{\lambda }{2\pi^2\alpha^4}\int_0^{+\infty} ds s^5\left\{f(\lambda,s)-\frac{\sqrt{\lambda}}{2(\lambda+s^2)^{1/2}}-\frac{\sqrt{\lambda}}{8(\lambda+s^2)^{5/2}}+\frac{5\sqrt{\lambda}}{16(\lambda+s^2)^{7/2}}\right\}.
\ee By means of a numerical evaluation, as $b_{11}\to 0$, we obtain
 \be
\langle T_{\mu}^{\mu}\rangle-\langle
T_{\mu}^{\mu}\rangle^{(0)}-\langle T_{\mu}^{\mu}\rangle^{(2)}\to
\frac{1}{240\pi^2\alpha^4}. \ee Therefore, the usual trace anomaly
is recovered in the limit $b_{11}\to 0$ only when the fourth
adiabatic order is also subtracted:
 \be
\langle T_{\mu}^{\mu}\rangle_{ren}= \langle
T_{\mu}^{\mu}\rangle-\langle T_{\mu}^{\mu}\rangle^{(0)}-\langle
T_{\mu}^{\mu}\rangle^{(2)}- \langle T_{\mu}^{\mu}\rangle^{(4)}\to
-\frac{1}{240\pi^2\alpha^4}. \ee The behaviour of the renormalized
trace $T$ as a function of $\lambda=\alpha/\sqrt{2 b_{11}}$ is
shown in Fig.\ref{traza4d}, where we see that it approaches to its
anomalous value as the dispersion relation tends to the usual one.

\section{Selfconsistent de Sitter solutions}\label{sec:desitter}

As an application of the results presented in the previous
sections, here we will study the SEE in de Sitter space-time,
$C(\eta)=\alpha^2/\eta^2$, taking  into account quantum effects of
free scalar fields with a generalized dispersion relation (i.e.,
including the backreaction of the quantum fields on the spacetime
dynamics).

It is well known that in de Sitter space-time the expectation
value of the energy momentum tensor $\langle T_{\mu\nu}\rangle$ of
a free scalar field with the usual dispersion relation is
proportional to the metric $g_{\mu\nu}$ in the de Sitter invariant
states \cite{Allen}. As they are related by  Bogoliubov
transformations with momentum independent coefficients, these
states form a one complex parameter family. Here, we will point
out that for a generalized dispersion relation there also exists a
one complex parameter family of states for which $\langle
T_{\mu\nu}\rangle\propto g_{\mu\nu}$, which is just the tensorial
structure of the cosmological constant term. We will then analyze
the relation between the de Sitter curvature ($R=12\alpha^{-2}$)
and the cosmological constant ($\Lambda$) for a dispersion
relation of the form $\omega_k^2=k^2+2b_{11}k^4/C(\eta)$.

In de Sitter space-time, the field modes $\chi_k$ satisfy (see
Eq.(\ref{PXXP})) \be \frac{\partial ^2\ck}{ \partial \eta^2}+\lp
\omega_k^2(\eta)+\frac{\tilde\mu^2 \alpha^2}{\eta^2}\rp\ck=0, \ee
where $\tilde\mu^2=m^2+n(n-1)(\xi-\xi_n)/\alpha^2$. With the
substitution $s= k\eta/\sqrt{\lambda}$, this equation can be
recast as \be \frac{\partial ^2\ck}{ \partial s^2}+\left(
\bar{\omega}^{2}(s)+\frac{\tilde\mu^2\alpha^2}{s^2}\right)\ck=0.
\label{eqdeSitter}\ee where
$\bar{\omega}^{2}(s)=\omega_k^{2}(\eta)/k^2$ is a function of $k$
and $\eta$ only through the variable $s$. Let $f(s)$ and $g(s)$ be
two independent solutions of this equation. Then, a field mode can
be conveniently expressed as \be \chi_k(s)=\frac{A_k}{\sqrt{k}}
f(s)+\frac{B_k}{\sqrt{k}} g(s). \ee We now choose the coefficients
$A_k=A$ and $B_k=B$ to be momentum independent. Thus, defining
$\psi(s)\equiv \sqrt{k}\chi_k(s)$, the normalization condition
(\ref{nor}) becomes \be \psi(s)\frac{\partial\psi^{*}}{\partial
s}(s)-\frac{\partial\psi}{\partial s}(s)\psi^{*}(s)=i
\sqrt{\lambda}. \ee By introducing these particular solutions
($\chi_k(s)=\psi(s)/\sqrt{k}$) into Eqs. (\ref{RHOO}) and
(\ref{PPP}), and rescaling the integration variable one can show
that $\rho =\langle T_{\eta\eta}\rangle /C$ and $p=\langle
T_{11}\rangle /C$ are time independent. Therefore, for the
corresponding states we have $\langle T_{\mu\nu}\rangle\propto
g_{\mu\nu}$, provided that $\langle T_{\mu\nu}\rangle$ is
covariantly conserved. It is remarkable that this one-parameter
family of states exists for any dispersion relation.

If we choose the particular state of the family that reproduces
the WKB solution as $|s|\to\infty$, the divergences in the stress
tensor can be absorbed into the bare gravitational constants, and
the SEE can be recast as \be G_{\mu\nu}+\Lambda g_{\mu\nu}=8\pi N
G_N \langle T_{\mu\nu}\rangle_{ren}, \ee where we $N$ is the
number of scalar fields and we assumed that the renomalized values
of $\alpha$, $\beta$ and $\gamma$ vanish. Using that in de Sitter
space-time $R_{\mu\nu}=R g_{\mu\nu}/4$ and $\langle
T_{\mu\nu}\rangle_{ren}=  T g_{\mu\nu}/4$ (where $T\equiv\langle
T^{\mu}_{\mu}\rangle_{ren}$), we have \be
-\frac{R}{4}+\Lambda=2\pi N G_N T.\ee
\begin{figure}[htp]
\includegraphics[width=8cm]{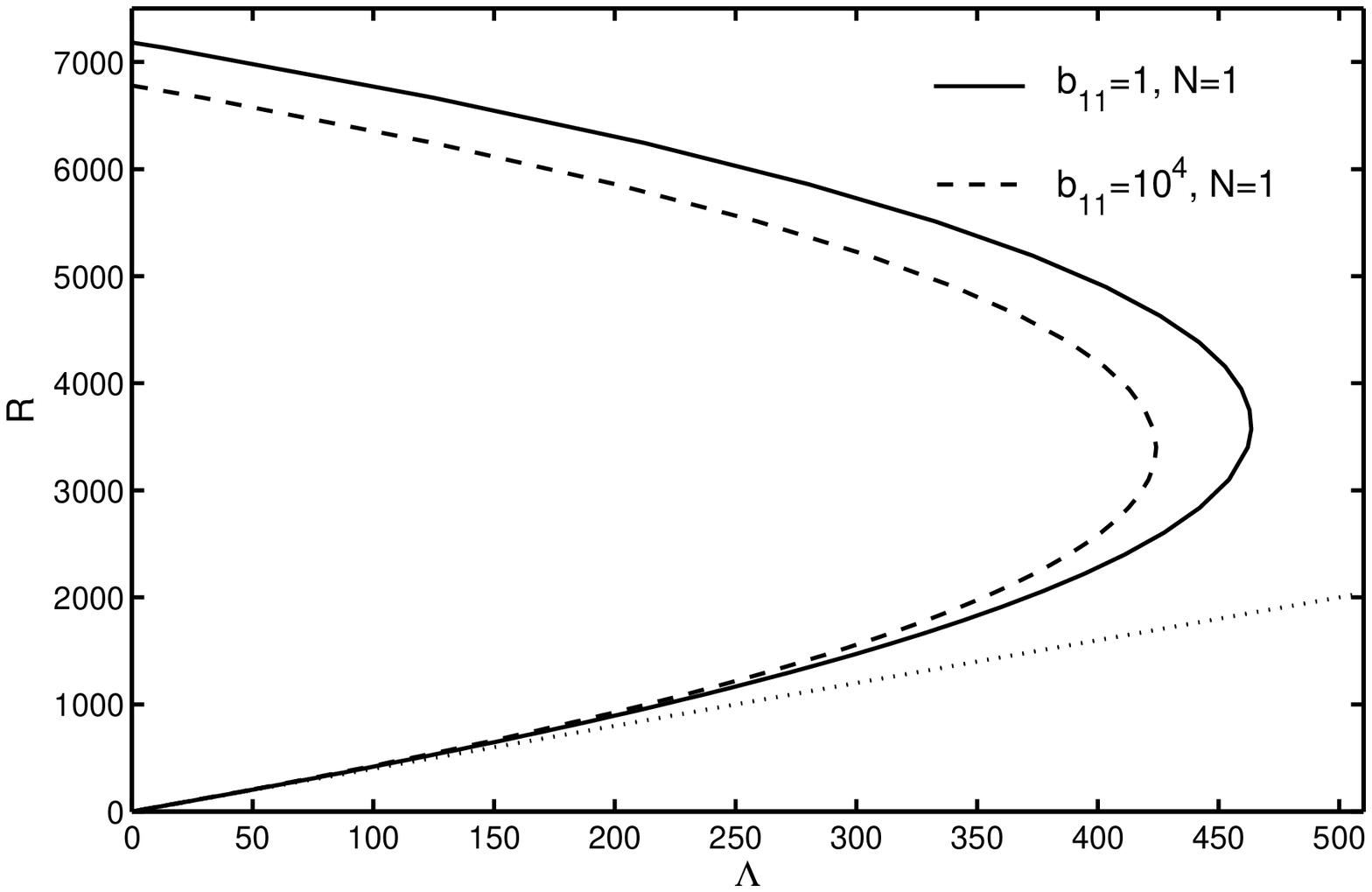}
\includegraphics[width=8cm]{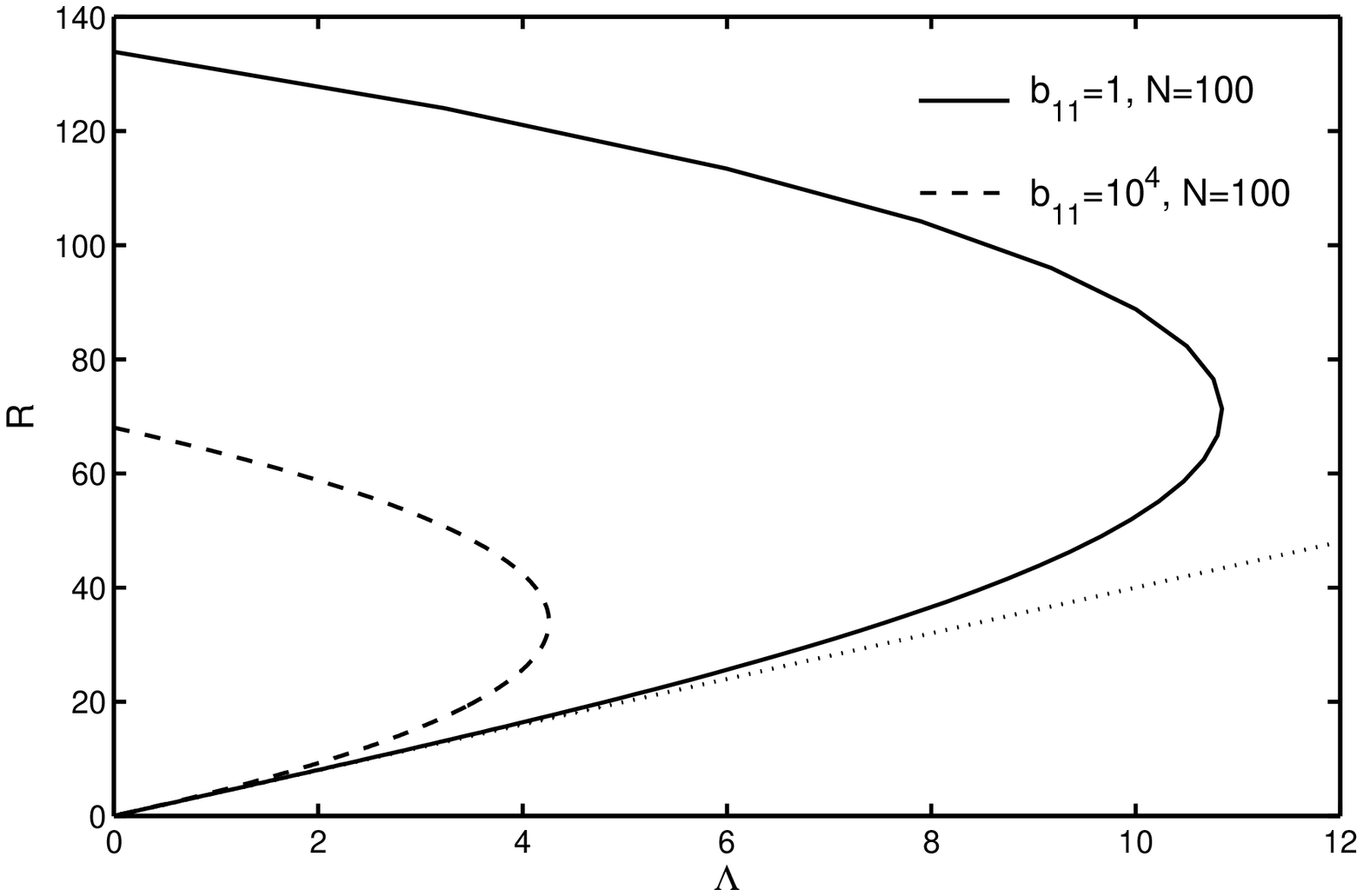}
\caption{The relation in Eq.(\ref{relation}) between the curvature
of the de Sitter space-time $R$ and the cosmological constant
$\Lambda$ for two values of $b_{11}$: In solid line $b_{11}=
m_{pl}^{-2}$ and in dashed line $b_{11}= 10^4 m_{pl}^{-2}$ (where
$m_{pl}$ is the Planck mass), with $N=1$ (on the left) and $N=100$
(on the right). In each case, the classical relation $\Lambda=R/4$
is shown by a dotted line. These results correspond to $N$
massless scalar fields with $\xi=1/6$ and
$\omega_k^2=k^2+2b_{11}k^4/C(\eta)$ in four dimensions. Note the
different scales in both axes.}\label{Rela}
\end{figure}Therefore, the de Sitter metric is a consistent solution of the SEE even when the
backreaction of scalar quantum fields with a generalized dispersion
relation is included.

For a massless field with $\xi=1/6$ and
$\omega_k^2=k^2+2b_{11}k^4/C(\eta)$ ($\tilde\mu^2=0$),
Eq.(\ref{eqdeSitter}) can be solved exactly. The solution for the
modes are given in Eq.(\ref{sol}) (where
$s=(2b_{11})^{1/4}\alpha^{-1/2}k\eta\equiv k\eta/\sqrt{\lambda}$).
Note that as $b_{11}\to 0$ these field modes tend to the standard
Bunch-Davies modes, defining a generalized Bunch-Davies vacuum. To
get the renormalized trace $T$ of the stress tensor we subtract up
to the fourth adiabatic order. An integral expression for the
unrenormalized trace with the subtraction of the zeroth and second
adiabatic order is given in Eq.(\ref{orden0y2}), while in
Eq.(\ref{orden4}) we have the fourth adiabatic order. Note that
the function $f\equiv T/R^2$ depends on only one free parameter
$b_{11} R=6/\lambda^{2}$. Therefore, we have \be
\Lambda(R)=\frac{R}{4}+2\pi N G_N T= \frac{R}{4}+2\pi N G_N R^2
f(b_{11}R).\label{relation}\ee We evaluate $f(b_{11}R)$
numerically for different values of $b_{11}R$. The relation in
Eq.(\ref{relation}) is shown in Fig. \ref{Rela}, where we have
also plotted the classical relation $\Lambda={R}/{4}$. The results
are presented for $b_{11}m_{pl}^2=1$ and $10^4$ (where $m_{pl}$ is
the Planck mass) for $N=1$ and $N=100$ (for intermediate values of
the parameters the results lie in between). From these figures, we
can see that for any of the values of the parameters the general
features are common. There is no selfconsistent solution for large
values of the cosmological constant $\Lambda$. The value of
$\Lambda$ above which there is no more selfconsistent solution
becomes larger as $b_{11}$ decreases. When $\Lambda$ is small
there are two selfconsistent solutions: one is near the classical
solution $\Lambda=R/4$ while the other has a positive curvature
$R$ larger than $4\Lambda$, even for negative values of $\Lambda$,
which in general is far from being in the semiclassical regime.
These results are similar to those obtained in Ref.\cite{Wada} for
the usual dispersion relation.

\section{An effective initial state?}

It has been suggested in the literature that the trans-Planckian
effects could be taken into account in a low energy effective
field theory with usual dispersion relations, by considering a
generic ``initial'' state for the modes of the quantum field when
they leave the sub-Planckian regime \cite{eff}. There is some
debate about whether the trans-Planckian corrections to the power
spectrum of primordial fluctuations can be consistently reproduced
from a suitable choice of the initial state
\cite{kalo,Padmanabhan,Holman}. Moreover, the choice of the
initial time for imposing the initial condition is a nontrivial
problem. In fact, as was pointed out in Ref.\cite{BrabderExSol},
an inadequate choice could lead to artificial oscillations in the
power spectrum.  On the other hand, within this description it is
difficult to quantify the backreaction effects \cite{transpbr}.

We will discuss now whether the backreaction effects on the
spacetime metric can be taken into account by considering an
arbitrary initial state in the low energy effective theory. If we
adopt the usual renormalization prescription in the effective
theory, the divergences in $\langle S|T_{\mu\nu}|S \rangle$ can be
absorbed into counterterms of the gravitational effective action
if the state $|S \rangle$ coincides with the adiabatic state up to
the fourth order (see for instance \cite{mpc, MolinaParis}).
Hence, the $\beta_k$ coefficient of the Bogoliubov transformation
that relate the mode function $\psi_k^S$ (corresponding to the
state $|S \rangle$) to the Bunch-Davies one $\psi_k^{BD}$, \be
\psi_k^S=\alpha_k\psi_k^{BD}+\beta_k\psi_k^{BD*}, \ee is required
vanish faster than $k^{-2}$ as $k\to\infty$. In previous works
\cite{kalo,Padmanabhan,MolinaParis} it has been shown that the de
Sitter invariant states belong to a one parameter family, which is
related to the Bunch-Davies vacuum by constant Bogoliubov
coefficients. Therefore, the only renormalizable state which is de
Sitter invariant is the Bunch-Davies vacuum.

In Section \ref{sec:desitter}, we have pointed out that for any
generalized dispersion relation, a family of states for which
$\langle T_{\mu\nu}\rangle\propto g_{\mu\nu}$ also exists (as for
the de Sitter invariant states in the usual theory). Thus, by
choosing the member of the family that tends to the adiabatic mode
of positive frequency for $|s|\to\infty$, one can obtain a
generalized Bunch-Davies state for which the renormalized stress
tensor is proportional to the metric tensor. Therefore, there are
states for which the expectation value of the stress tensor of a
scalar field with a generalized dispersion relation cannot be
reproduced by any choice of the initial quantum state of a field
with the standard dispersion relation, unless other
renormalization scheme is employed \cite{Holman}. In other words,
while can obtain selfconsistent de Sitter solutions for any
generalized dispersion relation, in the standard theory this is
possible only for a single quantum state.

%It has been suggested that the backreaction effects could
%not be negligible if the dispersion relation is such that there is a period
%where the WKB approximation
%becomes not valid for describing the evolution of the modes in the
%trans-Planckian regime \cite{adUnchanged}. However, in view of the
%above observations and the results of the previous section, if one
%choose a generalized Bunch-Davies vacuum, the backreaction of the
%scalar field fluctuations would only cause a finite shift in the
%background cosmological constant (in agreement with the suggestion
%of Ref.\cite{transpbr}). Moreover, the trans-Planckian
%corrections to the power spectrum represented by that state are
%expected to be irrelevant due to they are scale-invariant in a fixed de Sitter background.

\section{Final remarks}

In this paper we have presented a complete analysis of the
renormalization procedure for the semiclassical Einstein
equations, in theories in which the quantum scalar fields satisfy
modified dispersion relations. This work generalizes the adiabatic
renormalization for quantum field theory in curved spaces
developed in the seventies to theories containing higher spatial
derivatives of the matter fields.

We have shown that, even though power counting suggests that for
this class of theories it would be enough to renormalize the
cosmological and Newton's constants, in $3+1$ dimensions a
consistent procedure also involves the subtraction of the fourth
adiabatic order. Therefore, it is also necessary to include terms
quadratic in the curvature into the gravitational action in order
to absorb the divergences of the expectation value of the energy
momentum tensor. This subtle point was missed in our previous work
\cite{Nos}, and clarified in the $1+1$ dimensional case in
Ref.\cite{NosProc}.

We obtained regularized expressions for the stress tensor up to
the fourth adiabatic order, and showed explicitly the geometric
nature of its divergences. We also computed the trace of the
renormalized tensor in de Sitter space for the `would be conformal
field' in the standard theory, and recovered the usual trace
anomaly in the limit $M_c\to\infty$.

We have also shown that, whatever the dispersion relation, there
exist a family of quantum states for which the mean value of the
stress tensor is proportional to the metric. One member of this
family (the `generalized Bunch-Davies' vacuum), is renormalizable,
and therefore it allows the existence of selfconsistent de Sitter
solutions, as explicitly shown for massless, conformally coupled
fields. These solutions are not present for arbitrary states in
the standard theory (they only exist for the usual Bunch-Davies
vacuum), and therefore this particular trans-Planckian effect can
not be simulated by modifying the quantum state in an effective
theory with the usual dispersion relation.

\section*{Appendix A: Regularization of $\langle T_{\mu\nu}\rangle^{(4)}$}

In this Appendix we outline the procedure by which we derived  Eq.(\ref{Tmunuad4}), starting from the unrenormalized fourth adiabatic order of the stress tensor. The idea is to find
relations between the different integrals that appear in the WKB expansion of the stress tensor,
in order to reveal its geometric nature. The relations can be found by performing integrations
by parts and by using that in dimensional regularization the integral of a total derivative vanishes \cite{Collins}.

>From Eqs. (\ref{rhoad}) and (\ref{pad}), with the use of the WKB expansion, the fourth adiabatic order of  $\langle T_{\eta\eta}\rangle$ and $\langle T_{11}\rangle$ can be recast as
\begin{eqnarray}
\nonumber\langle T_{\eta\eta}\rangle^{(4)} &=&  \frac{\Omega_{n-1}\mu^{\bar{n}-n}}{
4 C (2\pi)^{(n-1)}}\int_{0}^{+\infty} \frac{dx\,x^{(n-3)/2}}{8 \,\tilde{\omega}_k^3}\left\{[{}^{(2)}W_k^2]^2-{}^{(2)}{W_k^2}'\mathcal{H}\left[\frac{f}{2}+2(n-1)(\xi-\xi_n)\right]\right.\\
&-&\left.{}^{(2)}W_{k}^2\mathcal{H}^2\left[\frac{5}{8}f^2+(n-1)(\xi-\xi_n)\left(3f-\frac{(n-2)}{2}\right)\right]\right\},
\label{rhoad4}
\end{eqnarray}
\begin{eqnarray}
\nonumber\langle T_{11}\rangle^{(4)} &=&  \frac{\Omega_{n-1}\mu^{\bar{n}-n}}{
4 C (2\pi)^{(n-1)}}\int_{0}^{+\infty} \frac{dx\,x^{(n-3)/2}}{8 \, \tilde{\omega}_k^3(n-1)}\left\{{{}^{(2)}W_k^2}'' [f+4(n-1)(\xi-\xi_n)]-[{}^{(2)}{W_k^2}]^2(2n-5-3f) \right.\nonumber\\
&+&{{}^{(2)}W_{k}^2}'\mathcal{H}\left[2(n-1)(\xi-\xi_n)(6f-(n-1))-\frac{f}{2}+\frac{5}{2}f^2\right]+{}^{(2)}W_k^2\mathcal{H}'\left[6(\xi-\xi_n)(n-1)f+\frac{(n-2)}{2}f+f^2
\right]\nonumber\\
&+&{}^{(2)}W_k^2\mathcal{H}^2\left[(\xi-\xi_n)(n-1)\left(\frac{1}{2}(n-1)(n-2)+9 f^2-6\dot{f}-3(n-1) f\right)+\frac{3}{2}f^3+\frac{(n-7)}{8}f^2\nonumber\right.\\
&-&\left.\left.f\dot{f}-\frac{(n-2)}{2}\dot{f}\right]\right\},
\end{eqnarray}where the function $f=f(x)$ is defined in Eq.(\ref{f}). Here we have used Eq.(\ref{W2ad4}) to write ${}^{(4)}W_k^2$ in terms of ${}^{(2)}W_k^2$ and its derivatives.
By using in addition that
\begin{subequations}
\begin{align}
{{}^{(2)}W^2_k}'&=\frac{\mathcal{H}''}{4}[4(\xi-\xi_n)(n-1)+f]+\frac{\mathcal{H}\mathcal{H}'}{8}[f^2-6\dot{f}+4(n-2)(n-1)(\xi-\xi_n)]+\frac{\mathcal{H}^3}{8}[2\ddot{f}-f \dot{f}],\\
{{}^{(2)}W^2_k}''&=\frac{{\mathcal{H}'}^2}{8}[f^2-6\dot{f}+4(n-2)(n-1)(\xi-\xi_n)]+\frac{\mathcal{H}\mathcal{H}''}{8}[f^2-8\dot{f}+4(n-2)(n-1)(\xi-\xi_n)]\\
\nonumber &+\frac{\mathcal{H}^4}{8}[\dot{f}^2+f \ddot{f}-2\dddot{f}]+\frac{\mathcal{H}^2\mathcal{H}'}{8}[12\ddot{f}-5f\dot{f}]+\frac{\mathcal{H}'''}{4}[4(\xi-\xi_n)(n-1)+f],
\end{align}
\end{subequations} we write $\langle T_{\eta\eta}\rangle^{(4)}$ and $\langle T_{11}\rangle^{(4)}$ in the form given in Eq.(\ref{tcoeff}), where the coefficients $\alpha_i$ and $\beta_i$ can be expressed as a linear combination of integrals of the form
\be
J _{m n l s}\equiv \int_{0}^{\infty}dx\frac{x^{\frac{(n-3)}{2}}}{\tilde{\omega}_k^3} {f}^m\, {\dot{f}}^n\,{\ddot{f}}^l\, {\dddot{f}}^s.
\ee with $m,n,l,s$ integer numbers. For example, the coefficient $\alpha_2$ considered in the text (Eq.(\ref{alpha2})) is given by
\be \label{alpha2Res}
\alpha_2=-\frac{1}{4}(n-1)^2(\xi-\xi_n)^2 I_3-\frac{1}{8} (n-1)(\xi-\xi_n) J_{1000}-\frac{1}{64}J_{2000},
\ee where $I_3=J_{0000}$ is defined in Table \ref{tabla}, and from Eq.(\ref{J12000}) we identify
\begin{subequations}\label{J00012re}
\begin{align}
J_{1000}&=\frac{(n-4)}{3}I_3,\label{J1000}\\
J_{2000}&=\frac{1}{15}(n-4)(n-6)I_3+\frac{2}{5}I_4.\label{J2000}
\end{align}
\end{subequations}
It is straightforward to show that
\be\label{relalpha4}\alpha_4=-\frac{1}{2}\alpha_2.\ee
To find a relation between $\alpha_1$ and $\alpha_2$ requires a little more work.
The coefficient $\alpha_1$ is given by
\be \label{alpha1}\alpha_1 = -\frac{3}{8}(n-1)^2(\xi-\xi_n)^2 J_{1000}-\frac{1}{16}(n-1)(\xi-\xi_n)[3 J_{2000}-2 J_{0100}]-\frac{1}{128}[3 J_{3000}-4 J_{1100}].\ee In order to express it in terms of the integrals $I_i$ in Table \ref{tabla} it is useful to observe that
\begin{eqnarray}
J_{2000}&=&\int_0^{\infty} \frac{x^{\frac{(n-3)}{2}}}{\tilde{\omega}_k^3} f \left(\frac{x}{\tilde{\omega}_k^2}\frac{d\tilde{\omega}_k^2}{dx}-1\right)\\\nonumber
&=& -J_{1000}+2\int_0^{\infty}\frac{x^{\frac{(n-1)}{2}}}{\tilde{\omega}_k^4} \frac{d\tilde{\omega}_k}{dx} f\\\nonumber
&=& \frac{(n-4)}{3}J_{1000}+\frac{2}{3} J_{0100},
\end{eqnarray} where the last equality follows after performing an integration by parts and discarding the surface term. Thus we have
\be \label{comb1} 3 J_{2000}-2 J_{0100}=(n-4) J_{1000},\ee which is one of the combination of integrals appearing in Eq.(\ref{alpha1}). In a similar way, the integral $J_{3000}$ can be recast as
\begin{eqnarray}
J_{3000}&=&\int_0^{\infty}\frac{x^{\frac{(n-3)}{2}}}{\tilde{\omega}_k^3} \left(\frac{x}{\tilde{\omega}_k^2}\frac{d\tilde{\omega}_k^2}{dx}-1\right) f^2\\\nonumber
&=&-J_{2000}+2\int_0^{\infty}\frac{x^{\frac{(n-1)}{2}}}{\tilde{\omega}_k^4} \frac{d\tilde{\omega}_k}{dx} f^2\\\nonumber
&=&-J_{2000}+\frac{(n-1)}{3}J_{2000}+\frac{4}{3} J_{1100},
\end{eqnarray} from which we have
\be\label{comb2} 3 J_{3000}-4 J_{1100}=(n-4) J_{2000}.\ee
Inserting Eqs. (\ref{J1000}), (\ref{comb1}) and (\ref{comb2}) into Eq.(\ref{alpha1}), we find
\be \alpha_1=-\frac{1}{8}(n-1)^2(\xi-\xi_n)^2(n-4) I_3-\frac{1}{16}(n-1)(\xi-\xi_n)(n-4) J_{1000}-\frac{1}{128} (n-4) J_{2000},\ee
and comparing this coefficient with $\alpha_2$ (given in Eq.(\ref{alpha2Res})) we see that
\be \label{relalpha1} \alpha_1=\frac{(n-4)}{2}\alpha_2. \ee
 Therefore, with the use of Eqs. (\ref{Hetaeta1}), (\ref{Hetaeta3}), (\ref{relalpha4}) and
 (\ref{relalpha1}), the component $\langle T_{\eta\eta}\rangle^{(4)}$ can be written as
 \begin{eqnarray} \label{tmunuetaeta}
 \langle T_{\eta\eta}\rangle^{(4)}&=&\frac{\Omega_{n-1}\,\mu^{\bar n-n}}{4(2\pi)^{n-1}} \left\{-\frac{\alpha_2 H_{\eta\eta}^{(1)}}{(n-1)^2} -\frac{32 H_{\eta\eta}^{(3)}}{(n-1)(n-2)(n-3)}\left[\alpha_3-\frac{(n-10)(n-2)}{32}\alpha_2\right] \right\}\\\nonumber
 &\equiv& B_1 H_{\eta\eta}^{(1)}+B_3 H_{\eta\eta}^{(3)}.
 \end{eqnarray} Since from Eqs. (\ref{alpha2}) and (\ref{J00012re}) the coefficient $\alpha_2$ can be expressed in terms of the integrals $I_i$, Eq.(\ref{coefB1}) for $B_1$ follows straightforwardly.

To show that the same equation (\ref{tmunuetaeta}) is satisfied by the component $\langle T_{11}\rangle^{(4)}$, with $H_{\eta\eta}^{(1,3)}$ replaced by $H_{11}^{(1,3)}$, we start by noting that
\begin{eqnarray}
  \beta_5&=&-\frac{2}{(n-1)}\alpha_2,\\
  \beta_2&=&\frac{(\xi-\xi_n)^2}{2}(n-1)\left[-\frac{1}{2}+(n-4)\right] I_3+\frac{(\xi-\xi_n)}{8}\left[6 J_{2000}-4 J_{0100}-J_{1000}\right]\\\nonumber
  &+&\frac{1}{64(n-1)}\left[6 J_{3000}-8 J_{1100}-J_{2000}\right],\\
   \beta_4&=&\frac{(\xi-\xi_n)^2}{8}(n-1)\left[9 J_{1000}+I_3\right] +\frac{(\xi-\xi_n)}{16}\left[9 J_{2000}-6 J_{0100}+J_{1000}\right]\\\nonumber
  &+&\frac{1}{128(n-1)}\left[9 J_{3000}-12 J_{1100}+J_{2000}\right].
   \end{eqnarray}
    With the use of Eqs. (\ref{J1000}), (\ref{comb1}) and (\ref{comb2}) we have
\begin{eqnarray}
  \beta_2&=&\left[-\frac{1}{2}+(n-4)\right]\beta_5, \\
  \beta_4&=&\left[\frac{1}{4}+\frac{3(n-4)}{4}\right]\beta_5.
   \end{eqnarray}
     Thus $\langle T_{11}\rangle^{(4)}$ can be expressed in the form
\begin{eqnarray} \label{tmunu11}
\langle T_{11}\rangle^{(4)}&=&\frac{\Omega_{n-1}\,\mu^{\bar n-n}}{4(2\pi)^{n-1}} \left\{-\frac{\alpha_2 H_{11}^{(1)}}{(n-1)^2} +\left[\beta_1-\left(\frac{3}{8}(n-4)(n-6)-\frac{3}{2}\right)\beta_5\right]\frac{\mathcal{H'}\mathcal{H}^2}{C} \right.\\\nonumber
 &+&\left.\left[\beta_3-\left(\frac{3}{16}+\frac{(n-4)}{64}(n^2-13 n+28)\right)\beta_5\right]\frac{\mathcal{H}^4}{C}\right\}\\\nonumber
 &\equiv& \left\{B_1 H_{11}^{(1)} +\tilde{\beta_1}\frac{\mathcal{H'}\mathcal{H}^2}{C} +\tilde{\beta_3}\frac{\mathcal{H}^4}{C}\right\}.
\end{eqnarray}
By the same procedure of integrating by parts and discarding surface terms,  and after a lot of  algebra, we find  \be
\tilde{\beta_3}=\frac{(n-5)}{8}\tilde{\beta_1}=\frac{(n-5)}{2 (n-2)(n-3)}B_3,
\ee where the coefficient $B_3$ is given in Eq.(\ref{coefB1}). Finally,  we arrive at Eq.(\ref{Tmunuad4}) by inserting these last relations into Eqs. (\ref{tmunuetaeta}) and (\ref{tmunu11}).

\begin{acknowledgments}
This work has been supported by  Universidad de Buenos Aires,
CONICET and ANPCyT. We would like to thank C. Simeone for useful
comments.

\end{acknowledgments}


\begin{thebibliography}{99}
\bibitem{signo} In Ref.\cite{Nos} there is a wrong sign in the last term of Eq.(\ref{PPP}).
\bibitem{f1} There is however a difference with the $1+1$ dimensional case:
in $3+1$ dimensions, with the usual dispersion relation, the
critical (fourth) adiabatic order is generally divergent, except
when $\xi=\xi_4=1/6$. Then the subtlety associated with the
renormalization prescription only appears for a generalized
dispersion relation. On the other hand, in $1+1$ dimensions the
second adiabatic order is convergent for any dispersion relation.
\end{thebibliography}
\end{document}